\documentclass[pra,twocolumn]{revtex4}

\usepackage{graphicx}

\begin{document}

\title{Polarized Fermi gases in asymmetric optical lattices}
\author{Xiaoling Cui and Yupeng Wang}
\affiliation{Beijing National Laboratory for Condensed Matter
Physics and Institute of Physics, Chinese Academy of Sciences,
Beijing 100190, China
\\}
\date{{\small \today}}
\begin{abstract}

The zero-temperature phase diagrams of imbalanced two-species Fermi
gases are investigated in asymmetric optical lattices with arbitrary
potential depths, based on the exact spectrum instead of the
Fermi-Hubbard model. We study the effect of lattice potentials and
atomic densities to the fully paired Bardeen-Cooper-Schrieffer (BCS)
state and particularly the Fulde-Ferrell-Larkin-Ovchinnikov (FFLO)
state. It is found that the increasing lattice potential favors BCS
at low densities because of the enhanced effective coupling; whereas
FFLO is favored at intermediate densities when the system undergoes
a dimensional crossover. Finally using local density approximation
we study the evolution of phase profile in the presence of external
harmonic traps by merely tuning the lattice potentials.
\end{abstract}

\maketitle

\section{Introduction}
Ever since the direct observation of phase separations of unequal
two-component $^6 Li$ cold atoms in
experiments\cite{zwierlein06,partridge06,zwierlein062,shin07}, the
topic of polarized Fermi gases has gained enormous interest from
theoreticians interests in the past few
years\cite{bedaque03,sheehy06,hu06,parish07np,wei07}. Such an
imbalanced and attractively interacting fermionic system makes it
possible to study in the laboratory several exotic fermionic pairing
mechanisms, including Sarma\cite{sarma63} or breached
pair\cite{liu03} state and Fulde-Ferrell-Larkin-Ovchinnikov (FFLO)
state\cite{fflo64,yoshida07} with finite pairing momentum. However,
it is hard for FFLO state to be experimentally observed in
three-dimensional (3D) free space. It was proved to survive in a
very small region in the Bardeen-Cooper-Schrieffer (BCS) side of
Feshbach resonance and totally vanish in the BEC side, both for a
homogeneous Fermi gas\cite{sheehy06,hu06,parish07np} and a trapped
one\cite{wei07}. Recently more and more attention has been paid to
one-dimensional (1D)\cite{huihu07,orso07} and two-dimensional
(2D)\cite{conduit08} systems. It has been pointed out that for these
reduced dimensions, the polarization window of FFLO state is
generally broadened compared to 3D free space.

Despite many previous studies on imbalanced fermions in pure 1D, 2D
and 3D systems, the crossover regimes between different dimensions
are poorly understood and still need further investigations, such as
quasi-1D and quasi-2D systems. Experimentally, these geometries can
be achieved by exposing magnetic or optical traps or optical
lattices along selective directions. Among all trapping
confinements, optical lattices are the most complicated because of
the band structures. A very interesting and important question then
immediately arises: How would the dimension and band structure
affect the fermionic superfluidity, especially the exotic FFLO
state? There have been several works related to this subject. For
example, the FFLO state was studied in pure 1D, 2D and 3D optical
latices and shown to be enhanced as a result of the Van Hove
singularities\cite{koponen07}; it was also previously found with
evident proportions in an array of 1D tubes produced by 2D optical
lattices\cite{parish07prl} and in two-lag ladders\cite{feiguin09}.
In the presence of optical lattices, however, all these studies have
adopted the Fermi-Hubbard model based on the tight-binding
approximation (TBA), which is generally believed to be valid for
deep lattices and not very strong interactions such that the lowest
band model is good enough to describe the system. On the other hand,
a different approach based on the exact lattice spectrum\cite{cui09}
shows that TBA does not necessarily produce the correct results
because of the multiband effects and deviations of the lowest band
structures, even for optical lattices with considerable band gap. In
this work we are going to use the exact spectrum to investigate the
zero temperature phase diagrams for imbalanced fermions in
asymmetric optical lattices, tuned by lattice potential depths and
atomic densities. This work studies the FFLO state in optical
lattices with arbitrary potential depths. The results obtained
reveal the fundamental effects of the dimension and the effective
coupling to the strength of fermionic superfluidities, including
FFLO state, and thus give an answer to the question previously
proposed. Therefore our results can serve as a guideline for cold
fermions experiments and might contribute to finally observing the
novel FFLO-type pairing state.

In this work we consider optical lattices applied in selective
directions and take into account the couplings between different
reciprocal lattice vectors, which recently were proved to be
necessary for optical lattices\cite{moon07,cui09}. The possibility
of FFLO pairing is studied in the remanent free direction(s). We
find that by increasing the lattice depths, FFLO gradually
diminishes at low atomic densities as a result of the enhanced
effective coupling, but revives at intermediate densities where the
system undergoes a dimensional crossover. Compared with the
density-driven BCS-BEC crossover initially studied in exciton
superconductors\cite{comte82} and then applied to attractive Fermi
gases\cite{NSR85}, the density-driven crossover from weak to strong
coupling limits mentioned here is unique to optical lattices and may
induce even more significant effects to the system in the sense that
it changes the effective dimension. In contrast, the strength of
unpolarized BCS superfluidity is enhanced at low atomic densities
but frustrated at intermediate ones because of the discontinuity of
the energy spectrum. All these results can be understood when
investigating the density of state (DOS), which has an intimate and
sensitive dependence on the lattice potential and the atomic
density. Finally, in the presence of the external harmonic trap, we
use local density approximation (LDA) to investigate the evolution
of phase profiles merely tuned by the potential depths of asymmetric
optical lattices.

The article is organized as follows. In Sec. \ref{model} we outline
our general model and method. In Sec. \ref{phasediag} we give the
phase diagrams for density-imbalanced two-component fermions in
several asymmetric optical lattices, including quasi-1D and quasi-2D
geometries. We mainly focus on quasi-1D case, which could be
achieved by applying optical lattices to either 3D or effective 2D
space. The effect of external harmonic confinement is studied in
Sec. \ref{h-trap}. We summarize our results in the last section.

\section{General Models}\label{model}

We consider a two-component Fermi gas with contact interactions by
the following Hamiltonian
\begin{eqnarray}
H&=&\int d\mathbf{r}\sum_{\sigma=\uparrow,\downarrow}
\psi^{\dag}_{\sigma}(\mathbf{r})\hat{H}_0(\mathbf{r})\psi_{\sigma}(\mathbf{r})+\nonumber\\
&&g \int d\mathbf{r}
\psi^{\dag}_{\uparrow}(\mathbf{r})\psi^{\dag}_{\downarrow}(\mathbf{r})\psi_{\downarrow}(\mathbf{r})\psi_{\uparrow}(\mathbf{r}),\label{hamil}
\end{eqnarray}
where $\hat{H}_0=\sum_{i=x,y,z}\hat{H}_i$ is composed by the kinetic
energy and optical lattice potential if applied; $g$ is the bare
contact interaction described by s-wave scattering length $a_s$ in
3D as $\frac{1}{g}=\frac{m}{4\pi a_s}-\frac{1}{V}\sum_{\mathbf{q}}
\frac{1}{2\epsilon_{\mathbf{q}}}$; and by the binding energy $E_B$
in effective 2D as $\frac{1}{g}=-\frac{1}{S}\sum_{\mathbf{q}}
\frac{1}{2\epsilon_{\mathbf{q}}+E_B}$. The renormalization of $g$
here is to eliminate the unphysical divergence due to the
high-momentum contribution in a Fermi gas.

To illustrate our scheme, we take a quasi-1D Fermi gas confined by
two orthogonal optical lattices in x and y directions. For example,
$\hat{H}_0=\sum_{i=x,y,z}-\partial_i^2/2m+V_0\big[\sin^2(\pi
x/a)+\sin^2(\pi y/a)\big]$, here $a$ is the lattice constant and
$V_0$ the potential depth. By mean-field treatment, first we expand
each field operator in terms of eigen-wavefunctions of $\hat{H}_0$,
$\psi_{\sigma}(\mathbf{r})=\sum_{\mathbf{nk}}\phi_{\mathbf{nk}}(\mathbf{r})
\psi_{\mathbf{nk}\sigma}$ with
\begin{equation}
\phi_{_{\mathbf{nk}}}(\mathbf{r})=\frac{1}{\sqrt{V}}e^{ik_{z}z}\sum_{\mathbf{G_{\bot}}}
a_{_{\mathbf{nk_{_{\bot}}}}}\hspace{-5pt}(\mathbf{G_{\bot}})e^{i(\mathbf{k_{\bot}}+\mathbf{G_{\bot}})\cdot\mathbf{r_{\bot}}},
\end{equation}
here
$\epsilon_{\mathbf{nk}}=\epsilon^0_{k_z}+\epsilon_{n_xk_x}+\epsilon_{n_yk_y}\
(\epsilon^0_{k_z}=\frac{k_z^2}{2m})$ is the eigen-energy;
$\mathbf{n}=\big\{n_x,n_y\big\}$ indicates the band index;
$\mathbf{k_{\bot}}$ lies in the first Brillouin Zone ($BZ$) while
$k_z$ has no constraint; $\mathbf{G_{\bot}}=2\pi/a(l_x,l_y)$ is the
reciprocal lattice vector. The Bloch wave functions and
eigen-energies can be obtained from the decoupled Schr\"odinger
equation in each (x or y) direction
\begin{eqnarray}
&&\sum_{G'}\Big\{\big[\frac{(k+G)^2}{2m}+\frac{V_0}{2}\big]\delta_{GG'}
-\frac{V_0}{4}\sum_i\delta_{G\pm\frac{2\pi}{a},G'}\Big\}\nonumber\\
&&\hspace{90pt} \times a_{nk}(G')=\epsilon_{nk}a_{nk}(G),
\end{eqnarray}
and the eigenvectors satisfy
$\sum_{G}a_{nk}^*(G)a_{n'k}(G)=\delta_{nn'}$ and
$a_{n,-k}(-G)=a_{nk}^*(G)$. For convenience we rescale the energy
and momentum respectively in units of the recoil energy
$E_R=\frac{1}{2m}(\frac{\pi}{a})^2$ and lattice reciprocal vector
$\frac{2\pi}{a}$. Then the lattice potential and atomic density are
expressed by two dimensionless parameters: $s=\frac{V_0}{E_R}$ and
$n=\frac{Na^3}{V}$.

Toward the FFLO state, we employ the simplest single plane wave
ansatz to the free direction as $\Delta(\mathbf{r})=\Delta_q
e^{iqz}$. For the lattice part, however, we focus on the most
probable pairing mechanism, i.e., pairing with two opposite crystal
momenta within the same band. The pairing between different
bands\cite{Martikainen08} is to be neglected here in our work. One
important reason is that we are dealing with the weak coupling
regime and therefore the interaction is not large enough to overcome
the energy differences and form inter-band pairs.

By employing two set of pairing fields in terms of Bloch state
indices $({\bf nk})$ and reciprocal lattice vectors $({\bf Q})$,
\begin{eqnarray}
\Delta_{\mathbf{Q}}&=&-\frac{g}{V}
\sum_{\mathbf{nk}}M^{\mathbf{Q}}_{\mathbf{nk}}\langle
\psi_{\mathbf{n, -k+\frac{q_z}{2}, \downarrow}}
\psi_{\mathbf{n,k+\frac{q_z}{2},\uparrow}}\rangle ,\nonumber\\
\Delta_{\mathbf{nk}}&=& \sum_{\mathbf{Q}}\Delta_\mathbf{Q}
M^{\mathbf{Q} \
*}_{\mathbf{nk}},\label{delt}
\end{eqnarray}
with $M^{\mathbf{Q}}_{\mathbf{nk}}=\sum_{\mathbf{G}}a_{\mathbf{n
-k}}(\mathbf{-G})a_{\mathbf{nk}}(\mathbf{G+Q})$ and $\mathbf{G},
\mathbf{Q}$ all lying in x-y plane, the Hamiltonian is then reduced
to
\begin{eqnarray}
&&H-\sum_{\sigma}\mu_{\sigma}N_{\sigma}=\sum_{\mathbf{nk}\sigma}(\epsilon_{\mathbf{nk}}-\mu_{\sigma})
\psi^{\dag}_{\mathbf{nk}\sigma}\psi_{\mathbf{nk}\sigma}-\nonumber\\
&&\sum_{\mathbf{nk}}(\Delta_{\mathbf{nk}}^*\psi_{\mathbf{n,
-k+\frac{q_z}{2}, \downarrow}}
\psi_{\mathbf{n,k+\frac{q_z}{2},\uparrow}}+h.c.)-\frac{V}{g}\sum_{\mathbf{Q}}
|\Delta_\mathbf{Q}|^2.\nonumber
\end{eqnarray}

By diagonalizing the Hamiltonian, the thermodynamic potential is
calculated at zero temperature as
\begin{eqnarray}
\frac{\Omega}{V}&=&\frac{1}{V}\sum_{\mathbf{nk}}\Big\{\Theta(-E_{\mathbf{nk}+})E_{\mathbf{nk}+}+
\Theta(-E_{\mathbf{nk}-})E_{\mathbf{nk}-}+\nonumber\\
&&\xi^{+}_{\mathbf{nk}}-\sqrt{\xi^{+\ \
2}_{\mathbf{nk}}+\Delta_{\mathbf{nk}}^{\ \
2}}\Big\}-\sum_{\mathbf{Q}}
\frac{|\Delta_{\mathbf{Q}}|^2}{g},\label{omega}
\end{eqnarray}
where the quasi-particle spectrum reads
\begin{equation}
E_{\mathbf{nk}\pm}=\sqrt{\xi^{+\ \
2}_{\mathbf{nk}}+\Delta_{\mathbf{nk}}^{\ \ 2}} \pm
\xi^{-}_{\mathbf{nk}}
\end{equation}
with $\xi^{+}_{\mathbf{nk}}=\epsilon_{n_x k_x}+\epsilon_{n_y
k_y}+(\epsilon^0_{k_z+q_z/2}+\epsilon^0_{-k_z+q_z/2})/2-\mu$,
$\xi^{-}_{\mathbf{nk}}=(\epsilon^0_{k_z+q_z/2}-\epsilon^0_{-k_z+q_z/2})/2-h$,
and chemical potentials $\mu=(\mu_{\uparrow}+\mu_{\downarrow})/2$,
$h=(\mu_{\uparrow}-\mu_{\downarrow})/2$.

From the saddle-point equations
$\partial\Omega/\partial\Delta_{\mathbf{Q}}^*=0$,
$\partial\Omega/\partial q_z=0$ and
$N_{\sigma}=-\partial\Omega/\partial\mu_{\sigma}$, we obtain the
gap, current and density equations as
\begin{eqnarray}
-\frac{\Delta_{\mathbf{Q}}}{g}&=&\frac{1}{V}\sum_{E_{\mathbf{nk}\pm}>0}\frac{M_{\mathbf{nk}}^{\mathbf{Q}}
\Delta_{\mathbf{nk}}}{2\sqrt{\xi^{+\ \
2}_{\mathbf{nk}}+\Delta_{\mathbf{nk}}^{\ \
2}}},\label{gap-eq}\\
n
q_z/2&+&\frac{1}{V}\big(\sum_{E_{\mathbf{nk}+}<0}-\sum_{E_{\mathbf{nk}-}<0}\big)k_z=0,\\
n&=&\frac{1}{V}\big(\sum_{\mathbf{nk}}1-\sum_{E_{\mathbf{nk}\pm}>0}
\frac{\xi^{+}_{\mathbf{nk}}}{\sqrt{\xi^{+\ \
2}_{\mathbf{nk}}+\Delta_{\mathbf{nk}}^{\ \
2}}}\big),\nonumber\\
\delta
n&=&\frac{1}{V}\big(\sum_{E_{\mathbf{nk}+}<0}1-\sum_{E_{\mathbf{nk}-}<0}1\big).\label{den-eq}
\end{eqnarray}

Therefore the consideration of couplings between different
$\mathbf{Q}$ directly results in the coupled gap equations
(\ref{gap-eq}). To get solutions with great precision one must take
into account as many non-zero $\mathbf{Q}$ as possible, and we find
that the smaller ones dominate over the larger ones, especially in
free space one has $\Delta_{\mathbf{Q}}=\Delta_0
\delta_{\mathbf{Q}0}$. This could be understood when examining the
properties of $M^{Q}_{nk}$, which is closely related to the pairing
amplitude $\Delta_{\mathbf{Q}}$ for each $\mathbf{Q}$, for a simple
1D optical lattice. At very high energy levels ($n\gg 1$),
$M^{Q}_{nk}\approx\delta_{Q0}$ close to those in free space, so the
contributions mainly come from several low bands. Near the bottom of
the lowest band, the perturbation theory justified for shallow
lattices ($s\ll 1$) gives $M^{Q}_{nk}=1(Q=0),\ \frac{s}{8}(Q=\pm1),\
\frac{s^2}{256}(Q=\pm2)$ and zero for other high-order ones; whereas
the TBA for deep lattices ($s\gg 1$) gives a smooth Gaussian
distribution as $M^{Q}_{nk}=\exp(-s^{-1/2}Q^{2})$. For intermediate
$s$ it is numerically verified that smaller $Q$ always lead to
predominantly larger $M^{Q}_{nk}$ and therefore contribute the most
to the summations in Eq.(\ref{delt}) and (\ref{gap-eq}).

In view of the properties of non-zero $\mathbf{Q}-$pairing in this
case, besides $\mathbf{Q}=0$ we consider the other four smallest
nonzero ones: $(\pm1,0),(0,\pm1)$.  Because of the equivalence of x
and y directions, all these non-zero $\mathbf{Q}$ share the same
pairing amplitude $\Delta_1$. Therefore we get
$\Delta_{\mathbf{nk}}=\Delta_0+2\Delta_1(M_{\mathbf{nk}}^{(10)*}+M_{\mathbf{nk}}^{(01)*})$,
and the coupled gap equations are in terms of $\Delta_0$ and
$\Delta_1$.

Next we utilize the this model to study the phase diagrams of
imbalanced fermions, either in terms of chemical potentials
$(h,\mu)$ or in terms of particle densities $(\delta n,n)$. In the
$h - \mu$ phase diagram, the phase boundaries separating the fully
paired BCS state, FFLO state, and normal (N) state are determined as
follows. We solve Eq.(\ref{gap-eq}) for BCS state at given $\mu$ and
obtain the minimum excitation energy $E_{min}=\min\sqrt{\xi_{nk}^{+\
2}+\Delta_{nk}^{\ 2}}$ as the first step. Then the BCS-normal
(BCS-N) phase boundary ($h_{BN}$) is determined from the following
equation
\begin{equation}
\Omega_{BCS}(\mu,h_{BN},\Delta=\{\Delta_{0},\Delta_{1}\})=\Omega_{N}(\mu,h_{BN},\Delta=\{0\}).
\end{equation}
Note that this equation has a solution for $h_{BN}(<E_{min})$ only
when $ \Omega_{BCS}(h=E_{min})>\Omega_{N}(h=E_{min})$. Otherwise the
stable magnetized superfluid ($SF_M$) phase would interpolate
between BCS and normal together with two continuous phase boundaries
in between. Near the FFLO-N boundary, $\Omega$ can be expanded in
terms of small pairing amplitudes as
$\frac{\Omega}{V}=\sum_{\mathbf{Q,Q'}}
W_{\mathbf{Q,Q'}}\Delta_{\mathbf{Q}}^*\Delta_{\mathbf{Q'}}$ where
\begin{equation}
W_{\mathbf{Q,Q'}}=-\frac{\delta_{\mathbf{Q,Q'}}}{g}-\frac{1}{V}\sum_{E_{\mathbf{nk}\pm}>0}\frac{M_{\mathbf{nk}}^{\mathbf{Q}}
 M_{\mathbf{nk}}^{\mathbf{Q'}*}}{2|\epsilon_{\mathbf{nk}}-\mu|}.
\end{equation}
For particular pairing momentum ($q$), the upper limit of $h$ for
FFLO state ($h_c$) is determined by setting $|W|=0$, and in this
particular case $W$ is a $5\times5$ matrix. While the FFLO-N
boundaries ($h_{FN}$ and $q_{FN}$) are obtained by finding the
maximum of $h_c$ as a function of $q$. Under mean-field treatment,
the BCS-N phase transition is always found to be first-order, while
FFLO-N is second-order. Moreover, as the thermodynamic potentials of
FFLO and normal states are very close to each other near the
BCS-FFLO boundary ($h_{BF}$), which makes $h_{BF}$ very close to
$h_{BN}$, in the following we do not distinguish these two
boundaries.

After obtain the $h - \mu$ phase diagram, we directly convert it to
the polarization-density ($P - n$) phase diagram composed of phase
separation (PS) of BCS and N state, FFLO and N state. For example,
the BCS-N phase boundary of the former ($h_{BN}$) corresponds to the
PS-N boundary of the latter ($P_{PN}$); in other words, $P_{PN}$ is
just the polarization of a N state with chemical potentials ($\mu,
h_{BN}$), for the reason that the PS with maximum polarization just
represents the critical point when BCS state vanishes and N spreads
to the whole space. Similarly the FFLO-N boundary of the former
($h_{FN}$) corresponds to the same one of the latter ($P_{FN}$).

\section{phase diagrams}\label{phasediag}

The phase diagrams of imbalanced two-component fermions including
FFLO state have been studied previously in free
3D\cite{sheehy06,hu06,parish07np} and 2D\cite{conduit08} spaces
based on mean-field theory, and in a 1D\cite{huihu07,orso07} space
using Bethe ansatz technique. Remarkably in 1D case FFLO exhibits a
quite broad polarization window ranging from a rather small value to
unity in the weak coupling limit. In this section we study the
quasi-1D and quasi-2D geometries that can be achieved by applying
optical lattices in selective directions. Note that there are two
ways to generate the quasi-1D system. One is from the combination of
a tight harmonic confinement in one direction and an optical lattice
in another, which generate 1D tubes lying in a plane; the other is
from two orthogonal optical lattices, which induce 1D tubes in 2D
lattice sites. In the next subsections we illustrate these two
cases. Finally we briefly introduce the quasi-2D system generated by
a 1D optical lattice.

For realistic simulations, we apply the cutoff momentum well above
the Fermi momentum $\mathbf{k_F}$, which is defined for an
unpolarized normal Fermi gas with the same total atoms number as the
polarized one studied, to ensure accuracy. The attractive s-wave
interaction is fixed to be well within the weak coupling limit.

\subsection{Quasi-1D geometry in an effective 2D space}\label{2d}

For an effective 2D space with the axial freedom of motion frozen by
a tight harmonic confinement, the renormalized atom-atom interaction
is characterized by the two-body binding energy $E_B=\frac{C
w}{\pi}{\rm exp}\big(\sqrt{2\pi}\frac{l}{a_s}\big)$\cite{petrov01},
where $w$ is the confinement frequency, $l=\sqrt{\frac{1}{mw}}$ is
the characteristic confinement length and $C\approx0.915$. In such a
2D system, all phase boundaries could be analytically
obtained\cite{he08,conduit08} except for FFLO. We numerically
calculate the FFLO-N boundary and verify the previous
predictions\cite{shima94,combescot05} in weak coupling limit that
two Fermi surfaces just touch at the critical point, with a constant
pairing momentum amplitude $q=2\sqrt{mE_B}$. The two relevant
boundaries are
\begin{eqnarray}
h_{BN}&=&\left\{\begin{array}{l} (\sqrt2-1)\mu+\frac{E_B}{\sqrt2}, \
\
-\frac{E_B}{2}\leq\mu\leq\frac{\sqrt2+1}{2}E_B \\
\sqrt{E_B(\mu+\frac{E_B}{4})},\ \ \ \ \mu>\frac{\sqrt2+1}{2}E_B
\end{array}\right.\\
h_{FN}&=&\sqrt{E_B(2\mu-E_B)}. \ \ \ \ \ \ \mu\geq E_B
\end{eqnarray}

\begin{figure}[ht]
\includegraphics[height=7.5cm,width=7cm]{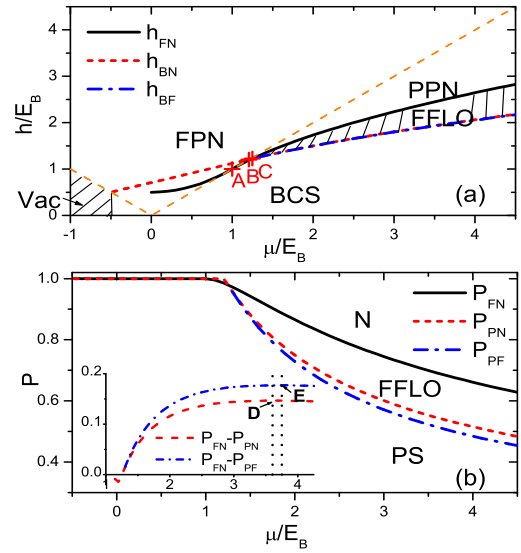}
\caption{(color online) Phase diagrams of a 2D Fermi gas. (a)$h-\mu$
diagram constructed by vacuum, BCS, FFLO, partially polarized normal
(PPN) and fully polarized normal (FPN) states. The boundary $h_{BF}$
is just slightly below $h_{BN}$. Red cross points $(\mu/E_B,h/E_B)$
from left to right: $A(1,1)$,
$B(\frac{\sqrt{2}+1}{2},\frac{\sqrt{2}+1}{2})$ and
$C(\frac{5}{4},\sqrt{\frac{3}{2}})$ correspond to $h_{FN}=\mu$,
$h_{BN}=\mu$ and $h_{FN}=h_{BN}$ respectively. Orange dashed line is
for $h=|\mu|$. (b)$P-\mu$ diagram constructed by phase separation
(PS), FFLO and normal (N) states. Inset shows two $\Delta P$ plots
$\Delta P_{1}=P_{FN}-P_{PN}$ and $\Delta P_{2}=P_{FN}-P_{PF}$, with
each peak $(\mu/E_B,\Delta P_{max})$ locating respectively at
$D(3.605,0.146)$ and $E(3.750,0.177)$. } \label{pure2d}
\end{figure}

The phase diagrams for a 2D Fermi gas are shown in Fig.
\ref{pure2d}, and here three points are emphasized in order. First,
numerical simulations show that the FFLO-N boundary $h_{FN}$, which
is the maximum value of $h_c$ at certain non-zero $q$, shows up
immediately when $\mu>0$, but is less than the BCS-N boundary
$h_{BN}$ until $\mu$ up to $\frac{5}{4}E_B$ at point C [shown in
Fig. \ref{pure2d}(a)]. Only beyond C, FFLO can exist as a candidate
of ground state. Second, Fig. \ref{pure2d}(b) shows that the maximum
polarization window for FFLO state to exist as ground state is less
than $20\%$, even when taking into account the small distinction
between $h_{BF}$ and $h_{BN}$. Third, throughout the range of $\mu$,
there is no stable magnetized superfluid ($SF_M$) phase found as
that in the strong coupling limit in 3D free space. One powerful
piece of evidence is that the Sarma-N boundary ($h_{SN}$) can never
exceed the lowest BCS excitation energy ($E_{min}$),
\begin{eqnarray}
h_{SN}&=&\left\{\begin{array}{l} E_B/2, \ \ \ \ \ \
|\mu|\leq E_B/2 \\
\sqrt{\mu E_B/2},\ \ \mu>E_B/2
\end{array}\right.\\
E_{min}&=&\left\{\begin{array}{l} \mu+E_B, \ \ \ \
-E_B/2<\mu\leq0 \\
\sqrt{E_B(2\mu+E_B)},\ \ \ \ \ \ \ \mu>0
\end{array}\right.
\end{eqnarray}
implying the exclusion of $SF_M$ from the ground-state solutions.
Here $h_{SN}$ is obtained by solving the gap equations at zero gap
amplitudes. We also use this criterion to judge the existence of
$SF_M$ in the following calculations for optical lattices.

\begin{figure}[ht]
\includegraphics[height=7.5cm,width=7cm]{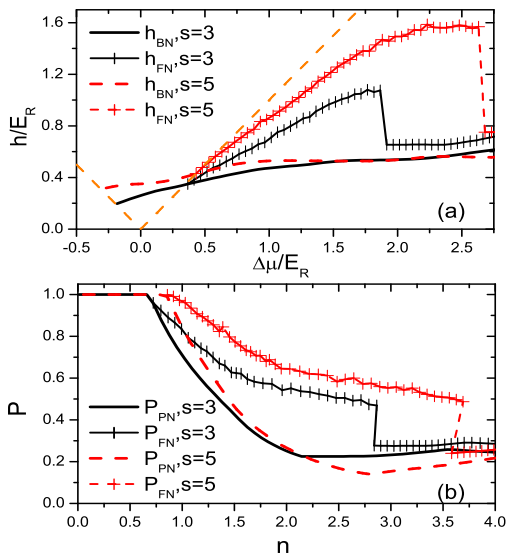}
\caption{(Color online) Phase diagrams of a 2D Fermi gas when
applying optical lattices along an arbitrary direction with
different depths $s=3,5$. The gas is in the weak interacting regime
with the two-body binding energy $E_B=0.2E_R$. (a)$h - \Delta\mu$
diagram with $\Delta \mu$ calculated from the bottom of the lowest
band. Orange dashed line is for $h=|\Delta \mu|$ defining the
threshold for a fully polarized normal state. BCS, FFLO and normal
states are separated by the corresponding boundaries as labeled in
the legend. FFLO exists as the ground state when $h_{BN}<h<h_{FN}$.
(b)$P - n$ diagram. From the bottom to top are phase seperation,
FFLO and normal state in turn. At a certain density, $P_{FN}$
suddenly drops to a lower value corresponding to the same behavior
of $h_{FN}$ in (a). } \label{q1d_m1}
\end{figure}
\begin{figure}[ht]
\includegraphics[height=7.5cm,width=7cm]{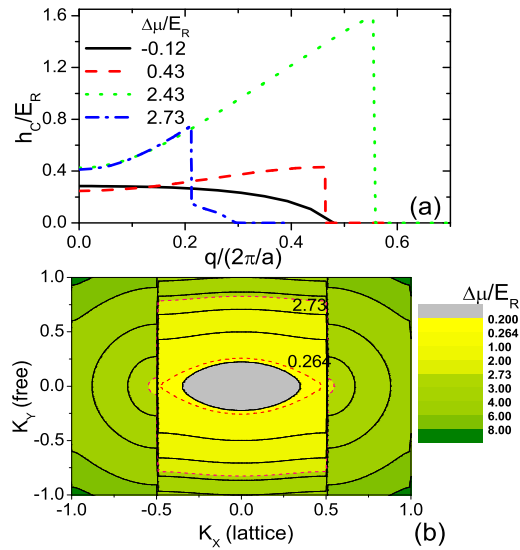}
\caption{(Color online) (a)Critical chemical potential differences
$(h_c)$ as a function of total pairing momenta $(q)$ at several
averaged chemical potentials ($\Delta \mu$) with lattice potential
$s=5$. (b)Fermi surface contour in momentum space. The optical
lattice is applied along x direction. The inner and outer red dashed
lines respectively represent the Fermi surfaces at the top of the
lowest band (labeled by 0.264) and slightly beyond the bottom of a
higher band (2.73). The outer line also corresponds to the
dashed-dotted curve in (a).} \label{q1d_m1_Hcq}
\end{figure}

When applying an optical lattice along one direction the phase
diagrams are dramatically changed as shown in Fig. \ref{q1d_m1}.
First we study the low density limit. It is observed that the
minimum density for FFLO to exist, $n_{min}$, moves to the right
(becomes large) as lattice potential $s$ increases. We argue that
this indicates a much more enhanced coupling of atoms moving in
deeper lattices. Actually this critical point ($n_{min}$) is an
analog of point C in Fig. \ref{pure2d}(a) for free 2D system, in
which $n_{min}$ monotonously increases with $\mu_C(=5E_B/4)$ and
also $E_B$. Optical lattices applied here drive the system to a
strongly interacting regime characterized by the enhanced effective
binding energies and thus increase $n_{min}$. Moreover, in this
dilute limit the maximum of critical chemical potential differences
$h_c(q)$ is always at $q=0$ (see Fig. \ref{q1d_m1_Hcq}(a)),
indicating no FFLO but only an unstable Sarma state
($h_{SN}<E_{min}$).

Second we observe that as the density $n$ or $\mu$ increases, the
FFLO-N boundary $h_{FN}$ initially goes up but then suddenly drops
at certain position, the reason for which is illustrated in Fig.
\ref{q1d_m1_Hcq}. As $n$ increases, $h_{FN}$ with nonzero pairing
momentum $q$ starts to exceed $h_{BN}$ which stabilizes FFLO states.
Meanwhile, the shape of Fermi surface distorts from an ellipse to
two disconnected lines, implying a crossover from 2D to a quasi-1D
geometry and correspondingly $h_{FN}$ has a rapid increase during
the crossover. When $n$ increases further, the shape of Fermi
surface changes again when atoms begin to fill a higher lattice
band. However, this occupation of a higher band exposes a
destructive effect on FFLO pairing along the free direction. For
example Fig. \ref{q1d_m1_Hcq}(a) shows that the maximum value of
$h_c$ for $\Delta\mu/E_R=2.73$ is much less than that for $2.43$,
leading to a sudden drop of $h_{FN}$ as well as $P_{FN}$ in Fig.
\ref{q1d_m1}. Moreover, it is noticed that the discontinuity of
$h_{FN}$ in this sensitive region produces double values of $P_{FN}$
at a definite density $n$ as a side effect, in which case the larger
one should be chosen as the real critical polarization for FFLO.
Finally it is shown by Fig. \ref{q1d_m1}(b) that at intermediate $n$
the PS-N boundary $P_{PN}$ is suppressed by increasing $s$, in
contrast to $P_{FN}$. We attribute this to the discontinuous lattice
spectrum that reduces the Hilbert space for pairing and therefore
disfavors the BCS superfluidity.

\begin{figure}[ht]
\includegraphics[height=4.5cm,width=7cm]{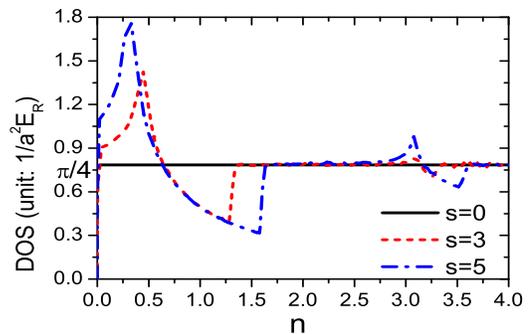}
\caption{(Color online) Density of state (DOS) scaled by
$1/(a^2E_R$) at the Fermi surface versus particle densities
$n=Na^2/S$. Optical lattices are applied along an arbitrary
direction in 2D space with different depths $s=0,3,5$. In free space
($s=0$), DOS is a constant ($\pi/4$).} \label{dos_q1d}
\end{figure}

Finally we comment that these effects of optical lattices at low and
intermediate densities can be well understood from the point of view
of density of state (DOS), see also Fig. \ref{dos_q1d}. The
variations of DOS bring two opposite effects to the BCS
superfluidity (SF) and FFLO. At low densities ($n$), DOS is enhanced
by the optical lattices applied and therefore favors SF but
disfavors FFLO as a result of the strong effective couplings; at
intermediate $n$, DOS decreases with $n$ which approximately
satisfies $DOS\sim 1/n$ indicating the evolution to 1D geometry, so
this time FFLO is favored while SF is suppressed. When $n$ further
increases, DOS becomes stable at the same level as in 2D free gas
when atoms start to fill a higher lattice band. After this point the
quasi-1D geometry fades away and FFLO is no longer favored.

\subsection{Quasi-1D and quasi-2D geometries in a 3D space}\label{3d_q1d}

\begin{figure}[ht]
\includegraphics[height=7.5cm,width=7cm]{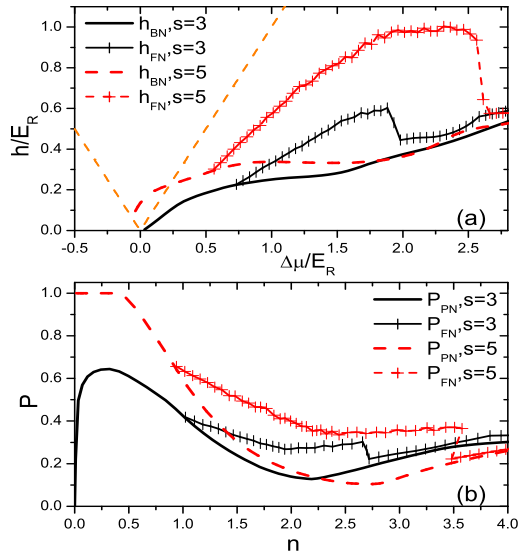}
\caption{(color online) Phase diagrams of a 3D Fermi gas when
applying optical lattices along two orthogonal directions with
different depths $s=3,5$. The s-wave scattering length is fixed to
be $a_s=-a/3$. (a)$h - \Delta\mu$ diagram. Orange dashed line is for
$h=|\Delta \mu|$. (b)$P - n$ diagram. Each boundary separates
corresponding phases in the same way as in Fig. \ref{q1d_m1}. }
\label{q1d_m2}
\end{figure}

The quasi-1D geometry in 3D space is formed  by applying optical
lattices along two orthogonal directions. The theoretical model has
been outlined in Sec. \ref{model}, and here we present the phase
diagrams in Fig. \ref{q1d_m2}.

At a given low density ($n$), the maximum polarization for phase
separation ($P_{PN}$) increases from a small value to unity with
lattice potential $s$, in contrast to quasi-1D geometry in 2D space;
however it does reveal the same fact that optical lattices enhance
the effective coupling. Moreover in this region, $h_{FN}<h_{BN}$ so
actually FFLO will not serve as a candidate for ground states. As
$n$ continues to increase, the system undergoes a crossover to
quasi-1D geometry, and $h_{FN}$ gradually goes up. Compared with a
negligible FFLO polarization window for the 3D free Fermi gas, the
optical lattice with $s=5$ can broaden the window from as large as
$25\%$ at the optimum density $n\simeq2.5$. When $n$ or $\mu$ is
large enough to touch a higher lattice band, $h_{FN}$ and $P_{FN}$
again suffer from a drop behavior as introduced in the previous
section.

Next we turn to quasi-2D geometry in 3D space which could be
generated by an axial optical lattice. We do not present the phase
diagrams here but give a brief introduction. Compared with quasi-1D
geometry the available FFLO window in this case is much less robust,
for instance less than $10\%$ with lattice potential $s=5$. Moreover
there is no obvious sudden drop of $h_{FN}$ or $P_{FN}$. This might
be ascribed to two tunable parameters ($|q|$ and $\theta$) in this
case for FFLO pairing momentum vector
$(q_x,q_y)=|q|(\cos\theta,\sin\theta)$, which make the pairing
relatively flexible and easily adaptive to external variations.

\section{Phase profile under external harmonic confinements}\label{h-trap}

In this section, we study the effect of asymmetric lattices to the
phase profile of imbalanced fermions in a harmonic trap using local
density approximation (LDA). In LDA, locally the gas is considered
to have the same properties as a bulk gas, which depend on the local
averaged chemical potentials
$\mu(\mathbf{r})=(\mu_{0\uparrow}+\mu_{0\downarrow})/2-V(\mathbf{r})$
and position-independent difference
$h=(\mu_{0\uparrow}-\mu_{0\downarrow})/2$. $\mu_{0\sigma}
(\sigma=\uparrow, \downarrow)$ is the chemical potential of
spin-$\sigma$ atoms at the trap center, and can be self-consistently
determined by the total particle numbers $N_{\sigma}$, s-wave
interactions and lattice potentials.

\begin{figure}[ht]
\includegraphics[height=7.5cm,width=8cm]{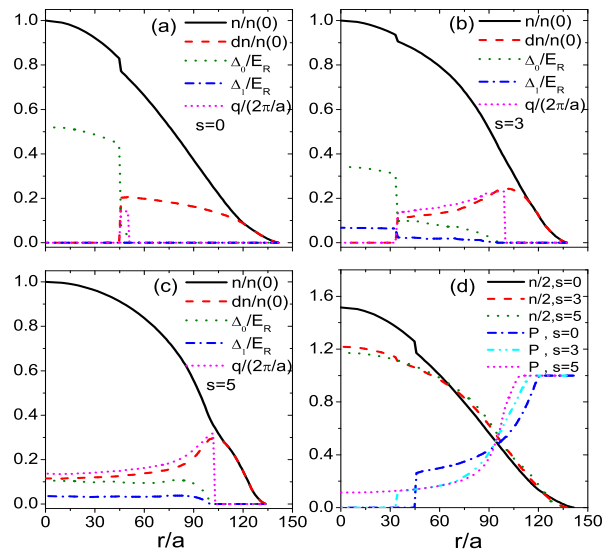}
\caption{(color online). (a,b,c) Spatial distributions of the
normalized total density ($n/n(0)$), density difference ($dn/n(0)$),
gap amplitudes ($\Delta_0, \Delta_1$), FFLO pairing momentum $q_z$
for a density-imbalanced two-species $^6 Li$ Fermi gas.
$n(0)=N(0)a^3/V$ is the dimensionless total density at the trap
center. The optical lattices $s=0(a), 3(b), 5(c)$ are applied along
$x$ and $y$ directions with lattice constant $a=532nm$. An isotropic
harmonic trap is also applied with trapping frequency $f=200Hz$. The
s-wave scattering length is fixed to be $a_s=-a/3$. Other parameters
are chosen so that these three plots (a,b,c) have nearly identical
particle numbers: $s, N/2, P, h/E_R, n(0)=0, 4.3(10^6), 0.42, 0.31,
3.03(a)$; $3, 4.35(10^6), 0.42, 0.32, 2.44(b)$; $5, 4.45(10^6),
0.43, 0.34, 2.35(c)$. (d) Spatial distributions of the averaged
total density ($n/2$) and polarization ($P$) corresponding to
(a-c).}\label{trap}
\end{figure}

In Fig. \ref{trap} we give three typical phase profiles under an
external isotropic harmonic trap. The optical lattices are applied
in two orthogonal directions and form an array of 1D tubes in 2D
lattice sites. Using LDA, we solve the ground state at each position
by minimizing the thermodynamic potential
$\Omega(\Delta_0,\Delta_1,q_z)$ in terms of its three parameters.
Meanwhile $\mu_{0\uparrow}$ and $\mu_{0\downarrow}$ are adjusted
such that the particle numbers are almost identical for different
$s$. Therefore Fig. \ref{trap} actually shows how the phase profile
of an imbalanced Fermi gas in a harmonic trap evolves when switching
on the optical lattices, especially for the FFLO state. Without
lattices ($s=0$), the profile is mostly constructed by BCS and
normal state, with negligible FFLO window in between. While
increasing $s$, the window is gradually broadened. For $s=3$, the
spatial range of FFLO has extended to be nearly as the same as that
of BCS and normal state. During this period, particles move from
center and edge to the middle region, causing both $n(0)$ (density
in the trap center) and $R_{TF}$ (Thomas-Fermi radius) to decrease
correspondingly. By increasing $s$ further, FFLO would gradually
spread to the trap center and take over BCS state, until finally
only FFLO and normal state are left in the trap, see Fig.
\ref{trap}(c) for $s=5$. Moreover, notice that the discontinuities
of densities $n_{\sigma}$, gap amplitudes $\Delta_0, \Delta_1$ at
the BCS-FFLO boundary imply a first-order phase transition, while
the continuities at FFLO-N boundary imply a continuous transition.

Finally two statements about FFLO are given as follows. First, FFLO
should be observed far away from the trap edge due to the atomic
dilution and strong effective interactions there; second, its
spatial range is enlarged by the increasing geometric asymmetry.

\section{summary}

In conclusion, we have studied the phase diagrams of imbalanced
two-species fermions in asymmetric optical lattices. We found that
the optical lattices applied expose two opposite effects on the BCS
and FFLO states, depending on the atomic densities. One is to
enhance the effective interaction at low densities which favors BCS
but destroys FFLO; the other is to induce the crossover to lower
dimension at intermediate fillings which favors FFLO, but destroys
BCS due to the discontinuity of lattice energy spectrum. Among all
the asymmetric geometries, we find that the quasi-1D system is the
most favorable one for observing the FFLO state. Finally by using
LDA, we present several typical phase profiles in a harmonic trap
with substantially different FFLO proportions, which are merely
tuned by the potential depths of applied asymmetric optical
lattices. These lattice effects still need to be further explored in
cold-atom experiments.

We are grateful to Wei Zhang for many stimulating discussions. This
work was financially supported by the National Science Foundating of
China, Chinese Academy of Sciences (CAS) and the 973-project of
Ministry of Science and Technology (MOST), China.

\end{document}